\documentclass[12pt,english]{article}
\usepackage[T1]{fontenc}
\usepackage[latin1]{inputenc}
\usepackage{subfigure}
\usepackage{float}
\usepackage{graphicx}
\usepackage{setspace}
\usepackage{amssymb}

\makeatletter

\AtBeginDocument{
  
}

\usepackage{babel}
\makeatother
\begin{document}
{\huge \noindent Graded and Binary Responses in Stochastic Gene Expression}{\huge \par}

{\Large \noindent Rajesh Karmakar and Indrani Bose$^{*}$}{\Large \par}

\noindent Department of Physics 

\noindent Bose Institute

\noindent 93/1, Acharya Prafulla Chandra Road, Kolkata-700 009, India.

\noindent $*$ Author to be contacted; e-mail: indrani@bosemain.boseinst.ac.in

\noindent \textbf{Abstract} 

\noindent {\small Recently, several theoretical and experimental
studies have been undertaken to probe the effect of stochasticity
on gene expression (GE). In experiments, the GE response to an inducing
signal in a cell, measured by the amount of $mRNAs/proteins$ synthesized,
is found to be either graded or binary. The latter type of response
gives rise to a bimodal distribution in protein levels in an ensemble
of cells. One possible origin of binary response is cellular bistability
achieved through positive feedback or autoregulation. In this paper,
we study a simple, stochastic model of GE and show that the origin
of binary response lies exclusively in stochasticity. The transitions
between the active and inactive states of the gene are random in nature.
Graded and binary responses occur in the model depending on the relative
stability of the activated and deactivated gene states with respect
to that of $mRNAs/proteins.$ The theoretical results on binary response
provide a good description of the {}``all-or-none'' phenomenon observed
in an eukaryotic system. }{\small \par}

\begin{singlespace}
\hfill{}
\end{singlespace}

\noindent \textbf{Keywords}: gene expression, graded and binary responses,
stochastic binary response, {}``all-or-none'' phenomenon, probability
density, activation

\begin{flushleft}\textbf{\Large 1. Introduction}\end{flushleft}{\Large \par}

\noindent Gene expression (GE) is the central activity in a living
cell. The two major steps in GE, transcription and translation, involve
several biochemical reactions. The time evolution of this system of
reactions or events is not a continuous process as molecular population
levels in a reacting system change only by discrete amounts. Furthermore,
the time evolution is not deterministic as the biochemical events
underlying GE are probabilistic in nature, i.e., the timing of the
biochemical events cannot be predicted with certainty. For example,
the binding/unbinding of $RNA$ polymerase $($$RNAP)$ at the promoter
region of $DNA$ and that of regulatory molecules at the operator
regions are probabilistic processes. The discrete, probabilistic nature
of the biochemical events may be ignored in the limit of large numbers
of participating biochemical molecules. In this case, the biochemical
reactions/events occur at much higher frequencies and fluctuations
around the mean levels of biomolecules participating in GE are small.
Thus, the time evolution of the system of reactions may be treated
to be continuous and deterministic as in the traditional differential
rate equation approach. In a living cell, the number of biomolecules
involved in GE is often small so that a stochastic rather than deterministic
description provides the more correct picture. In recent years, there
is a growing realization that stochasticity plays an important role
in determining the outcome of biochemical processes in the cell \cite{key-1,key-2}.
Stochastic effects in GE explain the pronounced cell-cell variation
observed in isogenic populations. A cell may have the option of proceeding
along one of two possible developmental pathways. The pathway selection
is probabilistic and the cell fate depends on the particular choice
of pathway. Thus, even a clonal population of cells can give rise
to two distinct subpopulations in the course of time. The randomization
of pathway choice leads to diversity and increases the likelihood
of survival of organisms in widely different environments. A well-known
example of the two way-choice is that of lysis-lysogeny in \emph{E.
Coli} \cite{key-3}.

The effect of stochasticity (randomness/noise) is prominent at the
level of an individual cell and can be masked due to ensemble averaging
in a population of cells. Single cell experiments provide evidence
that GE in a cell occurs in abrupt stochastic bursts \cite{key-4,key-5,key-6}.
In more recent experiments, a quantitative measure of noise associated
with GE has been obtained in both prokaryotic as well as eukaryotic
cells \cite{key-7,key-8,key-9}. A large number of theoretical studies
address the origin and consequences of stochasticity in GE \cite{key-3,key-10,key-11,key-12,key-13,key-14,key-15,key-16,key-17,key-18}.
Thus, the notion of stochasticity in GE is well established both theoretically
and experimentally.

Regulation of GE in a cell is achieved in a manifold of ways which
increase in complexity from prokaryotic to eukaryotic cells. In the
prokaryotic systems, regulation is achieved by the binding of regulatory
molecules (repressors or activators) to the operator regions of $DNA$.
In eukaryotes, the activator molecules are known as transcription
factors (TFs). Intra- and extra- cellular inducing signals activate
the TFs which then bind to appropriate enhancer sequences on the $DNA.$
The GE response to an inducing signal in an individual cell may be
graded or binary. Response is quantified by the amount of $mRNAs/proteins$
synthesized. In graded response, the output varies continuously as
the amount of input stimulus is varied till the steady state is reached.
In binary response, alternatively termed the {}``all-or-none'' phenomenon,
the output has a binary character, i.e., GE occurs at either a low
or a high level and expression at intermediate levels is rare. This
gives rise to a bimodal distribution in protein levels in an ensemble
of cells. Several experiments on both prokaryotic and eukaryotic cells
establish the binary character of GE \cite{key-6,key-9,key-19,key-20,key-21,key-22,key-23}.
Binary response may be ascribed to bistability which implies existence
of two stable steady states with low and high protein levels of GE.
One way of achieving bistability is through positive feedback or autocatalysis
in which the protein product of GE promotes further GE either directly
or via intermediates. The $lac$ operon in \emph{E. Coli} is an example
of a model system in which autocatalytic induction gives rise to the
{}``all-or-none'' phenomenon in GE \cite{key-19,key-23,key-24,key-25,key-26}.
Beckskei et al. \cite{key-23} have demonstrated that positive feedback
can generate binary response in a synthetic eukaryotic gene circuit.
In eukaryotic transcription, enhancers activate the usually weak eukaryotic
promoters. There is now strong experimental evidence that in some
systems enhancers do not affect transcription rate but rather increase
the probability of a gene synthesizing proteins at a high level \cite{key-27,key-28,key-29,key-30}.
In a population of cells, enhancers increase the number of cells expressing
at a high level but not the level of expression per cell.

Binary response in GE can have a purely stochastic origin. Kepler
and Elston \cite{key-10} provide examples of stochastic binary response
(SBR), i.e., binary response induced by noise. A simple model of SBR
shows a binary distribution of $mRNA$ levels in an ensemble of cells
\cite{key-31}. A recent model of eukaryotic GE suggests that fluctuations
in the binding of TFs to $DNA$ can explain graded and binary responses
observed in inducible GE \cite{key-32}. Fast chemical kinetics is
responsible for a graded response whereas slow kinetics leads to a
binary output. The {}``all-or-none'' phenomenon observed in some
eukaryotic systems does not involve positive feedback processes explicitly
\cite{key-4,key-5,key-6}. On the other hand, protein synthesis in
these systems occurs in stochastic bursts. Since the effect of stochasticity
is prominent in these systems, it is reasonable to conjecture that
the {}``all-or-none'' phenomenon (binary response) observed in these
systems is a manifestation of stochasticity. In this paper, we consider
a simple, stochastic model of GE studied earlier \cite{key-13,key-17,key-33}.
We show that graded and binary responses occur naturally in the model
depending on the relative stability of activated and deactivated gene
states with respect to that of $mRNAs/proteins$. Binary response,
obtained in the model, arises solely due to stochasticity and not
due to positive feedback processes. We further show that our model
gives a good description of the {}``all-or-none'' phenomenon observed
in an eukaryotic system \cite{key-6}.

\begin{flushleft}\textbf{\Large 2. Stochastic model of GE}\end{flushleft}{\Large \par}

\begin{flushleft}In the minimal model of GE, a gene can be in two
possible states: inactive ($G$) and active ($G^{*}$). Random transitions
occur between the states $G$ and $G^{*}$ according to the first
order kinetics\end{flushleft}

\begin{equation}
G\quad\begin{array}{c}
k_{a}\\
\rightleftharpoons\\
k_{d}\end{array}\quad G^{\star}\quad\begin{array}[b]{c}
j_{p}\\
\longrightarrow\end{array}\quad p\quad\begin{array}[b]{c}
k_{p}\\
\longrightarrow\end{array}\quad\Phi\label{mathed:first-eqn}\end{equation}

\noindent where $k_{a}$ and $k_{d}$ are the activation and deactivation
rate constants. In the active state $G^{*}$, transcription is initiated
followed by translation and protein synthesis. The separate processes
are combined into a single step of protein $(p)$ synthesis with the
rate constant $j_{p}$. The protein degrades with the rate constant
$k_{p}$ and the degradation product is represented as $\Phi$. If
cell division is taken into account, the protein decay rate has two
components, one the degradation rate and the other the dilution rate
of proteins due to cell growth and division. In this case, $k_{p}$
denotes the rate constant for protein decay.

In inducible GE systems, the activation of a gene is brought about
by an activator $S$, say, $TFs.$ The reaction scheme in the presence
of $S$ is given by 

\begin{equation}
G+S\quad\begin{array}{c}
k_{1}\\
\rightleftharpoons\\
k_{2}\end{array}\quad G_{-}S\quad\begin{array}{c}
k_{a}\\
\rightleftharpoons\\
k_{d}\end{array}\quad G^{\star}\quad\begin{array}[b]{c}
j_{p}\\
\longrightarrow\end{array}\quad p\quad\begin{array}[b]{c}
k_{p}\\
\longrightarrow\end{array}\quad\Phi\label{mathed:second-eqn}\end{equation}

\noindent where $G_{-}S$ represents the bound complex of $G$ and
$S$. The reaction scheme in equation (\ref{mathed:second-eqn}) can
be generalized by including direct transitions between $G$ and $G^{*}.$
The rate constants for transitions from $G$ to $G^{*}$ and $G^{*}$
to $G$ are $k_{on}$ and $k_{off}$ respectively. For eukaryotic
systems, the rate constant $k_{on}$ has a very low value as activating
TFs, $S,$ are required in most cases for transistion to the active
state $G^{*}.$ The reaction scheme is given by\[
\begin{array}{ccccc}
 & k_{1} &  & k_{a}\\
G+S & \rightleftharpoons & G_{-}S & \rightleftharpoons & G^{*}\\
 & k_{2} &  & k_{d}\end{array},\qquad\begin{array}{ccc}
 & k_{on}\\
G & \rightleftharpoons & G^{*}\\
 & k_{off}\end{array}\]
\begin{equation}
\begin{array}{ccccc}
 & j_{p} &  & k_{p}\\
G^{*} & \longrightarrow & p & \longrightarrow & \Phi\end{array}\label{mathed: eqn-3}\end{equation}

\noindent For eukaryotic systems, the initiation of transcription
by $RNA$ $polymerase$ II generally requires a prior assembly of
TFs on the enhancer regions of the target gene. This state of the
gene is represented by $G_{-}S$ in the reaction scheme \ref{mathed: eqn-3}.
The activating TFs, $S,$ facilitate the formation of the transcription
initiation complex which is bound to the promoter region of $DNA$
and consists of general TFs, other factors and $RNA$ $polymerase$
II. The gene is now in the active state $G^{*}$ and $RNAP$ starts
transcription after disengaging itself from the initiation complex
through the key step of phosphorylation. The general TFs are then
released allowing for the initiation of a new round of transcription
with another $RNAP$ molecule. In the simple reaction scheme \ref{mathed: eqn-3},
this is respresented by a return after transcription initiation to
the intermediate complex $G_{-}S$ and subsequent return to the active
state $G^{*}.$ 

If $n_{G}$ be the total concentration of genes then \begin{equation}
n_{G}=[G]+[G_{-}S]+[G^{*}]\label{mathed: eqn-4}\end{equation}

\noindent where $[G]$, $[G_{-}S]$, and $[G^{*}]$ denote the concentrations
of genes in the states $G$, $G_{-}S$, and $G^{*}$ respectively.
Using the method of King and Altman \cite{key-34}, the fractions
of genes in the inactive, intermediate and active states are given
by \[
\frac{[G]}{n_{G}}=\frac{k_{a}\: k_{off}+k_{2}\: k_{off}+k_{d}\: k_{2}}{A}\]
\[
\frac{[G_{-}S]}{n_{G}}=\frac{k_{d}\: k_{1}\: S+k_{on}\: k_{d}+k_{off}\: k_{1}\: S}{A}\]
\begin{equation}
\frac{[G^{*}]}{n_{G}}=\frac{k_{a}\: k_{1}\: S+k_{a\:}k_{on}+k_{2}\: k_{on}}{A}\label{mathed: eqn-5}\end{equation}

\begin{flushleft}respectively, where\begin{equation}
A=k_{a}\: k_{1}\: S+k_{a\:}k_{on}+k_{2}\: k_{on}+k_{d}\: k_{1}\: S+k_{on}\: k_{d}+k_{off}\: k_{1}\: S+k_{a}\: k_{off}+k_{2}\: k_{off}+k_{d}\: k_{2}\label{mathed: eqn-6}\end{equation}
From equation (\ref{mathed: eqn-5}), one can further write\end{flushleft}

\[
\begin{array}{ccc}
[G^{*}] & = & \frac{\frac{n_{G}\: k_{a}\:(\frac{S}{k_{s}}+\frac{1}{k})+k_{on}}{(1+\frac{1}{k}+\frac{S}{k_{s}})}}{\frac{k_{a}\:(\frac{S}{k_{s}}+\frac{1}{k})+k_{on}}{(1+\frac{1}{k}+\frac{S}{k_{s}})}+\{ k_{d}+\frac{k_{a}/k^{'}+k_{off}\,(1+\frac{S}{k_{s}})}{(1+\frac{1}{k}+\frac{S}{k_{s}})}\}}\end{array}\]

\begin{equation}
\begin{array}{cccccccccc}
\begin{array}{cccccccccc}
\begin{array}{ccccc}
\end{array} &  & = & \frac{n_{G}\: k_{a}^{'}}{k_{a}^{'}+k_{d}^{'}} & \qquad\qquad\end{array}\end{array}\label{mathed:eqn-7}\end{equation}

\noindent where \begin{equation}
k_{a}^{'}=\frac{k_{a}\:(\frac{S}{k_{s}}+\frac{1}{k})+k_{on}}{(1+\frac{1}{k}+\frac{S}{k_{s}})};\qquad k_{d}^{'}=k_{d}+\frac{k_{a}/k^{'}+k_{off}\,(1+\frac{S}{k_{s}})}{(1+\frac{1}{k}+\frac{S}{k_{s}})}\label{mathed: eqn-8}\end{equation}

\begin{flushleft}Also, \begin{equation}
k_{s}=\frac{k_{2}}{k_{1}},\qquad k=\frac{k_{2}}{k_{on}}\quad and\quad k^{'}=\frac{k_{2}}{k_{off}}\label{mathed:eqn-9}\end{equation}
\end{flushleft}

\noindent In the reaction scheme \ref{mathed:first-eqn}, the steady
state concentration of genes in the active state is given by\begin{equation}
[G^{*}]=\frac{n_{G}\: k_{a}}{k_{a}+k_{d}}\label{mathed: eqn-10}\end{equation}

\noindent Expressions (\ref{mathed:eqn-7}) and (\ref{mathed: eqn-10})
are identical in form with $k_{a}$ and $k_{d}$ replaced by $k_{a}^{'},$
$k_{d}^{'}.$ The equivalence relations in equation (\ref{mathed: eqn-8})
enable one to map the reaction scheme \ref{mathed: eqn-3} onto the
simpler scheme \ref{mathed:first-eqn} while calculating various quantities.
Use of the simpler reaction scheme leads to greater mathematical tractability.
The half-lives of the active and inactive states of the gene in the
reaction scheme \ref{mathed: eqn-3} are given by $T_{a}^{'}=log2/k_{a}^{'}$
and $T_{d}^{'}=log2/k_{d}^{'}$ respectively. Since $k_{a}^{'}$ and
$k_{d}^{'}$ are given by equation (\ref{mathed: eqn-8}), the half-lives
are dependent on $k_{a}$, $k_{d}$ as well as $S,$ the concentration
of TFs.

We now consider a simple stochastic model corresponding to reaction
scheme \ref{mathed:first-eqn}. The results we derive hold true for
the more complicated reaction scheme \ref{mathed: eqn-3} but with
$k_{a}$, $k_{d}$ replaced by $k_{a}^{'}$, $k_{d}^{'}$ (equation
(\ref{mathed: eqn-8})). At this point, one can ask about the validity
of the equivalence relations (equations (7) and (8)) in the stochastic
case. Use of the relations is justified only if the fluctuations in
the concentration $S$ of the activator molecules are ignored. Exact
validity can be established by deriving expressions for variance from
the Master Equations (treating $S$ to be constant) corresponding
to the reaction schemes 1 and 3. This has been done for the simpler
case $k_{on}=0$, $k_{off}=0$ (in the general case, these rate constants
are much smaller than the activation and deactivation rate constants
$k_{a}$ and $k_{d}$). The expressions for variance in the reaction
schemes 1 and 3, are found to be identical with $k_{a}$, $k_{d}$
in scheme 1 replaced by $k_{a}^{'}$, $k_{d}^{'}$ in scheme 3. In
the model, the only stochasticity arises from random transitions of
a gene between the inactive and active states as in the minimal model
of Cook et al. \cite{key-13}. Protein synthesis from the active gene
and protein degradation occur in a deterministic manner. In each state
of the gene, the concentration of proteins evolves deterministically
according to the equation\begin{equation}
\frac{dx}{dt}=\frac{j_{p}}{X_{max}}z\;-\; k_{p}\: x\:=\: f(x,z)\label{mathed: eqn-11}\end{equation}

\noindent where $z=1(0)$ when the gene is in the active (inactive)
state and $x=\frac{X}{X_{max}},$ $X$ and $X_{max}$ being the protein
concentration at time t and the maximum protein concentration respectively.
The variable $x$ thus denotes protein concentration normalized by
the maximum possible concentration. The latter quantity is equal to
the protein concentration in the steady state if the gene is always
in the active state, i.e, deactivation processes are disallowed. We
note that $X_{max}=\frac{j_{p}}{k_{p}}.$ Let $p_{j}(x,t)$ $(j=0,$
$1)$ be the probability density function when $z=j.$ The total probability
density function is \begin{equation}
p(x,t)=p_{0}(x,t)+p_{1}(x,t)\label{mathed: eqn-12}\end{equation}

\begin{flushleft}The rate of change of probability density is given
by\begin{equation}
\frac{\partial p_{j}(x,t)}{\partial t}=-\frac{\partial}{\partial x}[f(x,j)\: p_{j}(x,t)]+\sum_{k\neq j}[W_{kj}\: p_{k}(x,t)-W_{jk}\: p_{j}(x,t)]\label{mathed:eqn-13}\end{equation}
\end{flushleft}

\noindent where $W_{kj}$ is the transition rate from the state $k$
to the state $j$ and $W_{jk}$ is the same for the reverse transition.
The first term in equation (\ref{mathed:eqn-13}) is the so called
{}``transport'' term representing the net flow of the probability
density. The second term represents the gain/loss in the probability
density due to random transitions between the state $j$ and other
accessible states. In the present case, equation (\ref{mathed:eqn-13})
gives rise to the following two equations:\begin{equation}
\frac{\partial p_{0}(x,t)}{\partial t}=-\frac{\partial}{\partial x}(-k_{p}\: x\: p_{0}(x,t))+k_{d}\: p_{1}(x,t)-k_{a}\: p_{0}(x,t)\label{mathed: eqn-14}\end{equation}

\begin{equation}
\frac{\partial p_{1}(x,t)}{\partial t}=-\frac{\partial}{\partial x}\{(\frac{j_{p}}{X_{max}}-k_{p}\: x)\: p_{1}(x,t)\}+k_{a}\: p_{0}(x,t)-k_{d}\: p_{1}(x,t)\label{mathed: eqn-15}\end{equation}

\begin{flushleft}The steady state distribution ($\frac{\partial p_{0}(x,t)}{\partial t}=0,\:\frac{\partial p_{1}(x,t)}{\partial t}=0$)
is given by\begin{equation}
p(x)=N\: x^{(\frac{k_{a}}{k_{p}}-1)}(1-x)^{(\frac{k_{d}}{k_{p}}-1)}\label{mathed: eqn-16}\end{equation}
\end{flushleft}

\noindent where $N$, the normalization constant, is given by the
inverse of a beta function \cite{key-14}\begin{equation}
N=\frac{1}{B(\frac{k_{a}}{k_{p}},\:\frac{k_{d}}{k_{p}})}\label{mathed: eqn-17}\end{equation}

\begin{figure}
\subfigure[]{\includegraphics[%
  width=1.8in]{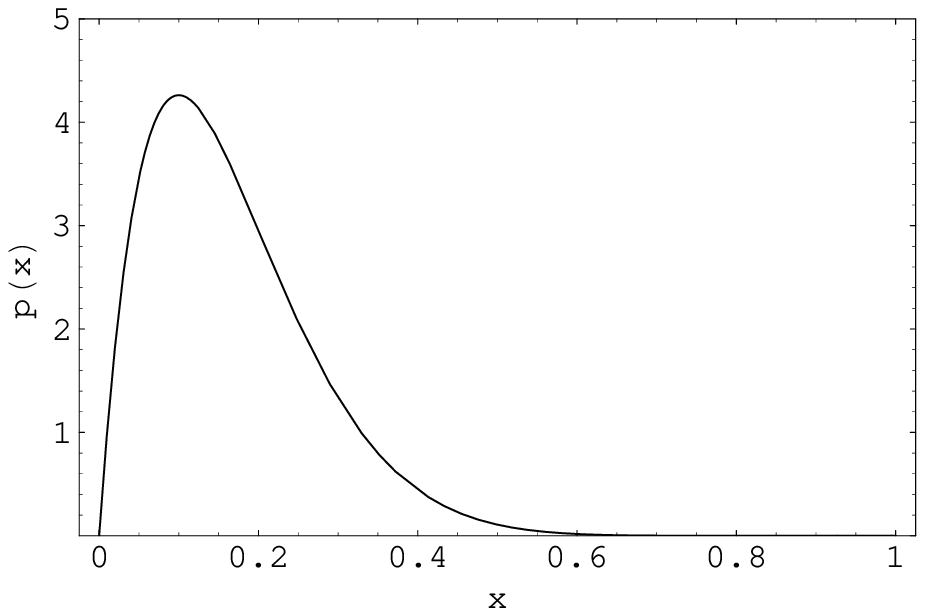}}\subfigure[]{\includegraphics[%
  width=1.8in]{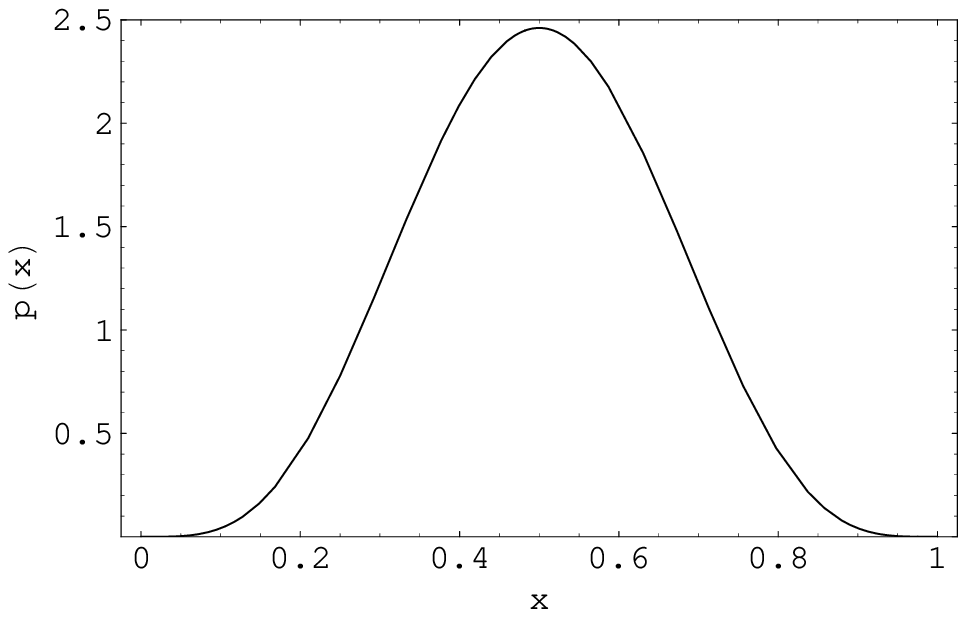}}\subfigure[]{\includegraphics[%
  width=1.8in]{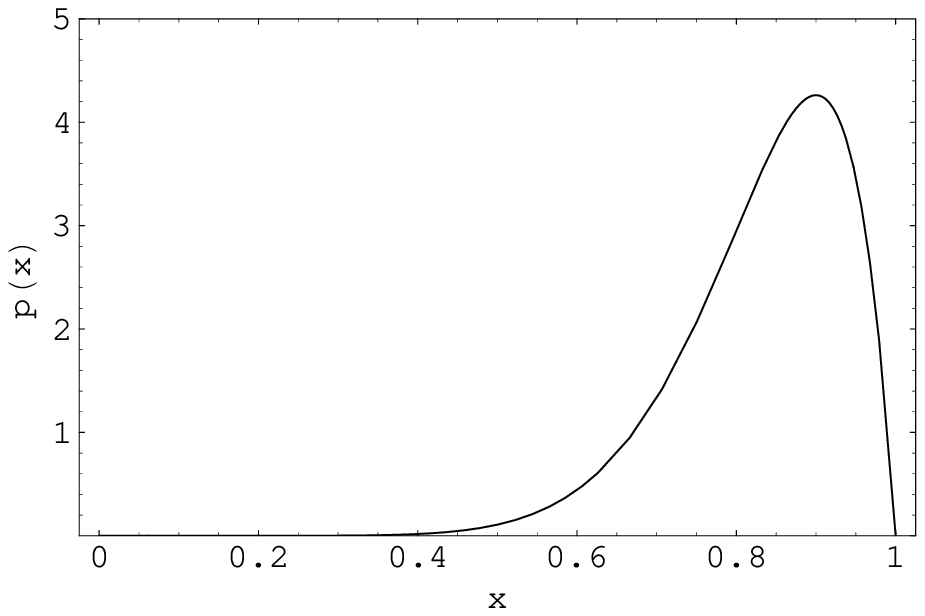}}

\caption{Plot of $p(x)$ versus $x$ (Case I, $r_{1},\: r_{2}>1$): 1A ($r_{2}>>r_{1})$,
1B ($r_{2}=r_{1})$ and 1C ($r_{1}>>r_{2})$ respectively.}
\end{figure}

\begin{flushleft}\textbf{\Large 3. Graded and binary responses} \textbf{}\end{flushleft}

\noindent The graded and binary responses to varying concentrations
of $S$ can be understood by considering two limiting cases of the
steady state distribution of protein levels $p(x)$ (equation (\ref{mathed: eqn-16})):
$\frac{k_{a}}{k_{p}},\;\frac{k_{d}}{k_{p}}>1$ (Case I) and $\frac{k_{a}}{k_{p}},\;\frac{k_{d}}{k_{p}}<1$
(Case II). Figure 1 shows plots of $p(x)$ versus $x$ corresponding
to Case I. The mean value of $x$ is given by\begin{equation}
<x>\;=\:\int_{0}^{1}\; x\; p(x)\; dx\label{mathed: eqn-18}\end{equation}

\begin{equation}
=\frac{j_{p}}{k_{p}}\frac{k_{a}}{k_{a}+k_{d}}\label{mathed: eqn-19}\end{equation}

\begin{flushleft}Define $r_{1}=\frac{k_{a}}{k_{p}}$ and $r_{2}=\frac{k_{d}}{k_{p}}.$
Figures 1A, 1B and 1C correspond to $r_{2}>>r_{1}>1,$ $r_{1}=r_{2}>1$
and $r_{1}>>r_{2}>1$ respectively. In the presence of an inducing
stimulus, $k_{a}$ and $k_{d}$ are replaced by $k_{a}^{'}$ and $k_{d}^{'}$
(equation (\ref{mathed: eqn-8})). As $r_{1}=\frac{k_{a}^{'}}{k_{p}}$
increases, the mean protein level increases from a lower to a higher
value. The increase in $r_{1}$ can be brought about by increasing
the concentration of $S$. Thus the mean protein level is a continuous
function of $[S],$ i.e., a graded response is obtained. Saturation
level is reached when $\frac{S}{k_{s}}$ in equation (\ref{mathed: eqn-8})
is $>>$ $1$ so that $k_{a}^{'}=k_{a}$ and $k_{d}^{'}=k_{d}+k_{off}$
. \end{flushleft}

Figure 2 shows plots of $p(x)$ versus $x$ corresponding to Case
II, i.e., $r_{1}<1,$ $r_{2}<1.$ In this case $p(x)$ is peaked at
a low (zero) value of $x$ ($r_{1}<<r_{2},$ figure 2A), a high value
of $x$ ($r_{1}>>r_{2},$ figure 2D) or at both low and high values
of $x$ (figures 2B and 2C). Thus, in a cell GE predominantly occurs
at low and/or high levels and protein production at intermediate levels
is negligible. Again, in the presence of an inducing stimulus, $S$,
$r_{1}=\frac{k_{a}^{'}}{k_{p}}$ and $r_{2}=\frac{k_{d}^{'}}{k_{p}}$
can be changed by changing the concentration $[S].$ The response in this
case is not graded as the mean protein level is not a continuous function
of $[S${]} but has only low and high values. Figures 2B and 2C correspond
to SBR and bifurcation from a unimodal probability distribution function
(figure 2A) to a bimodal one is brought about by varying $r_{1}$
and $r_{2}.$ SBR gives rise to the {}``all-or-none'' phenomenon
in GE. In experiments on a population of cells, a fraction of cells
is found to be in the state with low (zero) level of GE and another
fraction is in the state with high level of GE. The fraction of cells
in which protein synthesis occurs at intermediate levels is small.
In the cases when $r_{1}>1,\: r_{2}<1$ and $r_{1}<1,\: r_{2}>1$,
unimodal responses are obtained. In the first case, GE occurs at a
high level and in the second case, GE occurs at a low level. The response
is not graded in the presence of an inducing stimulus.

The graded and binary responses to an inducing stimulus are a natural
outcome of stochastic gene activation and deactivation processes.
If the gene is always in the inactive state ($z=0$ in equation (\ref{mathed: eqn-11})),
the mean protein level corresponds to $x=0.$ If the gene is in the
active state $(z=1)$ and no deactivation processes are allowed, the
mean protein level is given by $\frac{j_{p}}{k_{p}}$ and $x=1$ corresponding
to maximum protein synthesis. When stochastic GE is considered, i.e.,
random activation/deactivation processes are taken into account, two
possibilities arise. If the activation and deactivation rates are
faster than the protein degradation rate, an average protein level
is obtained due to the accumulation of proteins over random transitions
between the values $x=0$ and $x=1.$ In the opposite case, i.e.,
when the activation and deactivation rates are slower than the protein
degradation rate, the mean protein level is either $x=0$ or $x=1$
depending on whether the gene is in the inactive or the active state.
The half-life of each such state is larger than that of synthesized
proteins so that in each case sufficient time is available for the
mean protein level to attain its particular steady state value. Due
to the relatively larger protein degradation rate, there is no accumulation
of proteins over the random transitions so that observed protein levels
are predominantly at $x=0$ and $x=1.$ 

Ko \cite{key-11} has considered a stochastic model for gene induction
and has shown using computer simulation that two types of response
are possible. GE in the model is switched on and off due to the binding
and unbinding of the TF-complex at the gene. Stochasticity is introduced
into the model because of the probabilistic nature of the binding
and unbinding events. An unstable transcription complex causes a {}``homogeneous''
level of gene induction while a stable transcription complex gives
rise to a {}``heterogeneous'' level. The homogeneous case is analogous
to graded response and the binary response is an example of heterogeneous
response. In the detailed stochastic model studied by Pirone and Elston
\cite{key-32}, fluctuations in TF binding are shown to explain graded
and binary responses to an inducing stimulus. A binary pattern of
GE is obtained when the enhancer-state fluctuations (caused by the
binding and unbinding of TFs) are slow whereas faster enhancer-state
fluctuation give rise to a graded response. The conclusions are arrived
at by using a combination of approximate analytical methods and numerical
techniques like Monte Carlo (MC) simulation based on the Gillespie
Algorithm. The role of operator fluctuations in transcriptional regulation
has been studied by Kepler and Elston \cite{key-10} using the Master
Equation Approach. In the limit of large protein abundance, an equation
similar to equation (\ref{mathed:eqn-13}) is obtained. Again, the
interpretation is the same. In each state of the operator, the protein
concentration evolves deterministically but there are random transitions
between the two states of the operator corresponding to the occupation
and unoccupation of the operator region by an activator. The analysis
is not, however, extended further. 

\begin{figure}[H]
\subfigure[]{\includegraphics[%
  width=2.7in]{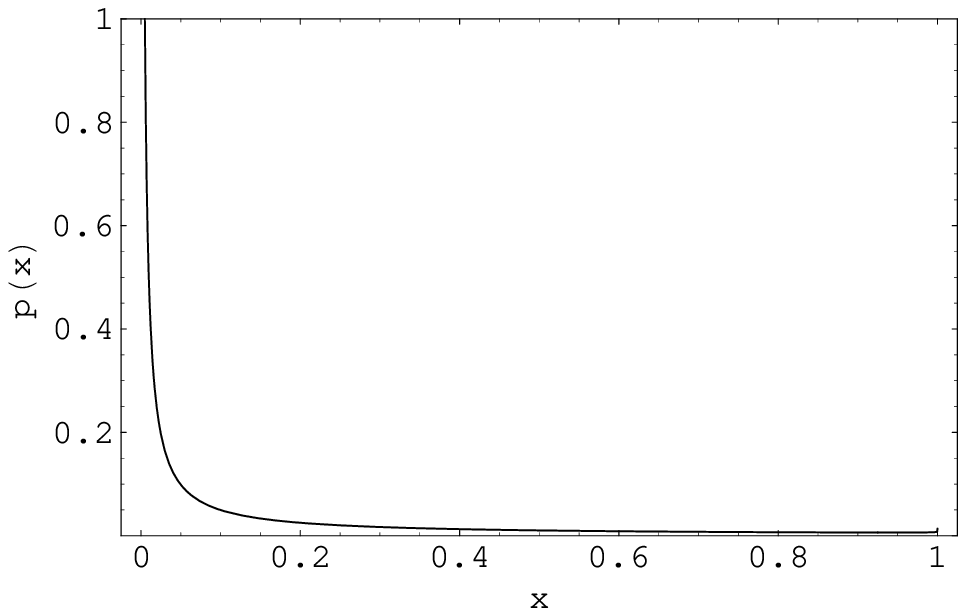}}\subfigure[]{\includegraphics[%
  width=2.7in]{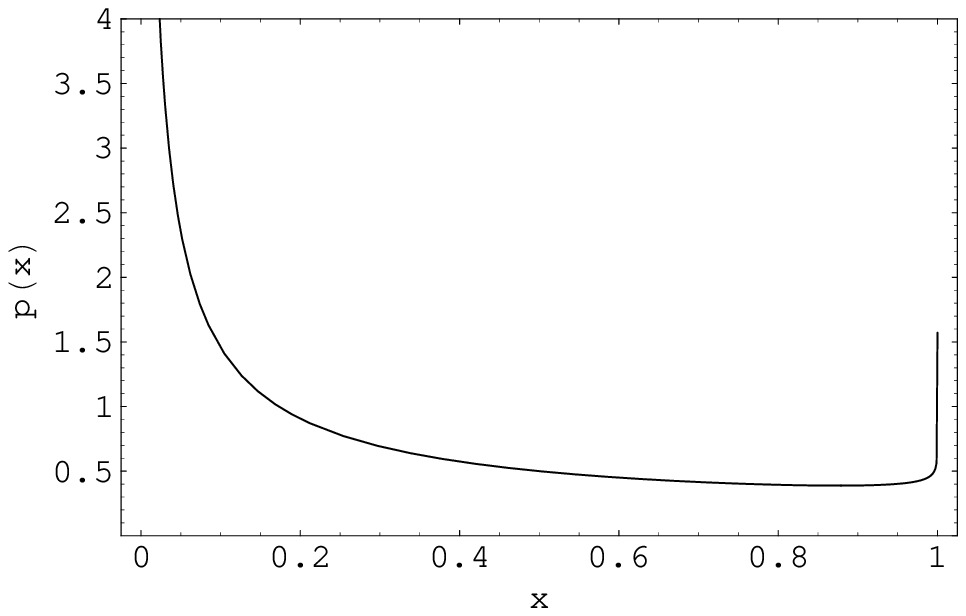}} \subfigure[]{\includegraphics[%
  width=2.7in]{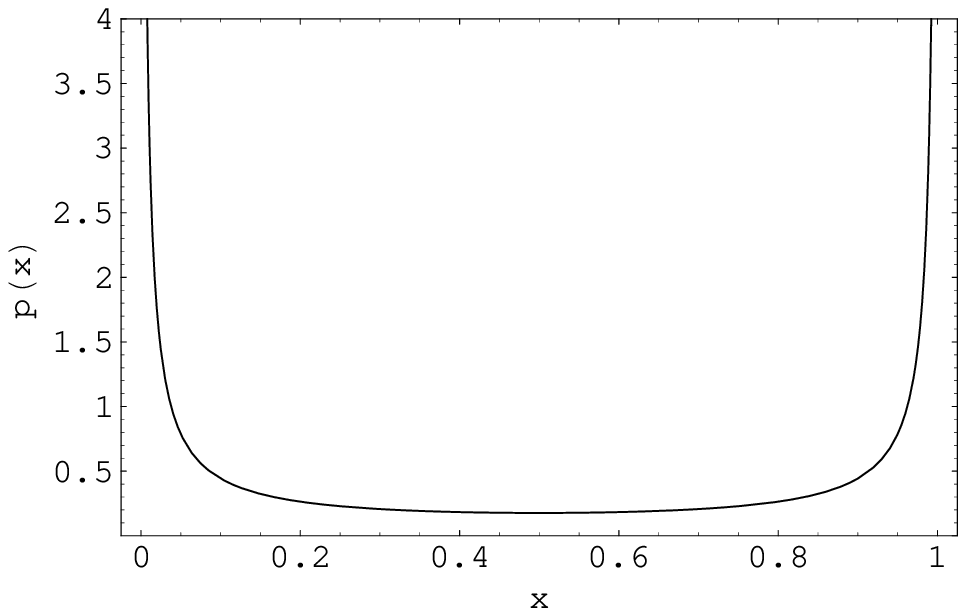}}\subfigure[]{\includegraphics[%
  width=2.7in]{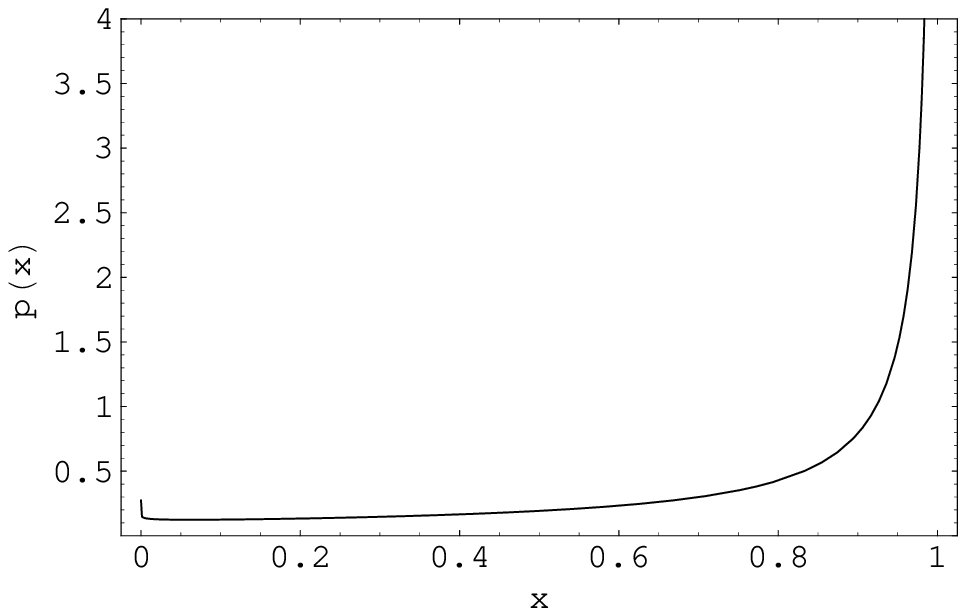}}

\caption{Plot of $p(x)$ versus $x$ (Case II, $r_{1},\: r_{2}<1$): 2A ($r_{2}>>r_{1}),$
2B $(r_{2}>r_{1}),$ 2C $($$r_{2}=r_{1})$, and 2D ($r_{1}>>r_{2})$
respectively. }
\end{figure}

Discovery of biologically active molecules, say, drugs involves testing
the effect of such molecules on appropriate targets. Membrane receptors
are the largest class of drug targets. Drugs interacting with receptors
are broadly of two types: antagonists and agonists. Antagonists block
receptor activity whereas agonists binding to the receptors enhance
cellular activity. The binding triggers a series of biochemical events
which lead to a change in cellular activity. The change can be linked
to the expression of a reporter gene so that detection and quantification
of the response to agonist-induced receptor activation are possible.
Figure 3 shows a cartoon of how the reporter gene conveys information
regarding receptor activation \cite{key-6}. A cascade of intracellular
processes are initiated by the binding of an agonist to the receptor.
This is accompanied by a change in the concentrations of messengers
in the cell. As a result, TFs are activated which then translocate
to the nucleus. The TFs bind to the target gene and initiate expression
of both this gene as well the reporter gene. The $mRNA$, generated
from the reporter gene, is translated into the reporter enzyme. The
reporter enzyme catalyzes the conversion of substrates into detectable
products. 

The scenario depicted in figure 3 provides the basis for high-throughput
screening of pharmaceutical candidate drugs in living mammalian cells
\cite{key-6}. The reporter gene used is $\beta-lactamase$ the protein
product of which hydrolyzes a substrate giving rise to a large shift
in fluorescence emission wavelength. Cells in which the reporter gene
is not expressed or expressed at a very low level appear fluorescent
green whereas reporter-positive cells with a high level of GE appear
fluorescent blue. The activation of cellular processes is brought
about by the binding of the agonist carbachol to the muscarinic receptor.
In the experiment, the percentages of blue, blue-green (intermediate
level of GE) and green cells are measured by flow cytometry with varying
carbachol dose and also as a function of time after stimulation by
carbachol. The major finding is that as the carbachol dose increases
from a low to high value, the fraction of green cells (low level of
GE) decreases and that of blue cells (high level of GE) increases.
The percentage of blue-green cells remains fairly low throughout.
This is a manifestation of the {}``all-or-none'' phenomenon, i.e.,
binary response in GE.

\begin{figure}
\begin{center}\includegraphics[%
  width=3in]{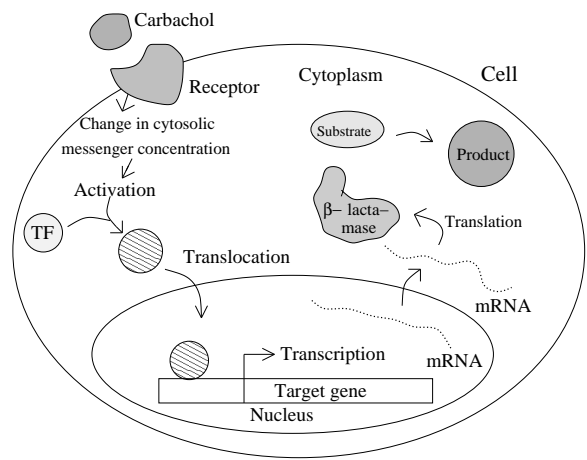}\end{center}

\caption{Schematic diagram showing the action of carbachol.}
\end{figure}

We now show that the simple stochastic model studied by us provides
a good quantitative fit to the experimental data of Zlokarnik et al.
Since the probability density function $p(x)$ of protein levels is
known (equation (\ref{mathed: eqn-16})), one can calculate experimentally
measurable quantities like the dose-response function. Figure 4 shows
the experimental data points corresponding to fractions of blue $+$
blue-green (depicted by solid triangles) and blue (depicted by solid
squares) cells versus $log\,(C)$, where $C$ is the carbachol concentration.
The fraction of blue-green cells is given by the difference in data
points belonging to the two curves. The remaining cell fractions describe
green cells. The concentration of activated TFs ($S$ in our model)
may be taken to be proportional to the concentration $C$ of carbachol
and in our theoretical dose-response curves (solid lines in figure
4), $[S]$ replaces $C.$ From equation (\ref{mathed: eqn-16}), the
steady state probability of finding a cell with $x$ (mean protein
level divided by maximum protein concentration), greater than a threshold
value $x_{thr}$ is\begin{equation}
p(x>x_{thr})=1-\frac{\int_{0}^{x_{thr}}\: x^{(\frac{k_{a}}{k_{p}}-1)}\:(1-x)^{(\frac{k_{d}}{k_{p}}-1)}\: dx}{\int_{0}^{1}\: x^{(\frac{k_{a}}{k_{p}}-1)}\:(1-x)^{(\frac{k_{d}}{k_{p}}-1)}\: dx}\label{mathed: eqn-20}\end{equation}
\begin{equation}
=1-\frac{_{k_{p}\: x_{thr}^{\frac{k_{a}}{k_{p}}}\: F_{1}[1-\frac{k_{d}}{k_{p}},\:\frac{k_{a}}{k_{p}},\:1+\frac{k_{a}}{k_{p}},\: x_{thr}]}}{k_{a}\: B(\frac{k_{a}}{k_{p}},\:\frac{k_{d}}{k_{p}})}\label{mathed: eqn-21}\end{equation}

\noindent where $_{2}F_{1}(a,b,c;z)$ is the hypergeometric function
\cite{key-35}. In our model, we assume that a cell is in a state
with high level of GE if the mean protein level in the steady state
is greater than a fraction of $0.9$ of the maximum protein concentration
i.e., $x>0.9.$ By setting $x_{thr}=0.9$ in equation (\ref{mathed: eqn-20})
and replacing $k_{a},$ $k_{d}$ by the effective rate constants $k_{a}^{'},$
$k_{d}^{'}$ (equation (\ref{mathed: eqn-8})), one can calculate
$p(x>x_{thr})$ for various values of $S.$ The probability $p(x>x_{thr})$
can also be interpreted as the fraction of cells in a cell population
with $x>x_{thr}$. The theoretical dose-response curve obtained in
this manner gives a good fit to the experimental data points (solid
squares in figure 4) for the parameter values (in arbitrary units)
$k_{2}=1.6\times10^{-4},$ $k_{on}=1.2\times10^{-6},$ $k_{off}=1.32\times10^{-4},$
$k_{s}=1.6\times10^{-6},$ $k_{p}=1,$ $k_{a}=0.17$ and $k_{d}=0.0465$.
The data points in this case correspond to the fraction of cells in
a high level of GE (blue cells). A cell is assumed to be in a state
with low level of GE if $x$ is $<$ $x_{thr}=0.1$ (green cells).
A cell is in a state with intermediate level of GE when $0.1<x<0.9$
(blue-green cells). The cell fraction in the last case is given by\begin{equation}
p(0.1<x<0.9)=\frac{\int_{0.1}^{0.9}\; x^{(\frac{k_{a}}{k_{p}}-1)}\;(1-x)^{(\frac{k_{d}}{k_{p}}-1)}\; dx}{\int_{0}^{1}\; x^{(\frac{k_{a}}{k_{p}}-1)}\;(1-x)^{(\frac{k_{d}}{k_{p}}-1)}\; dx}\label{mathed: eqn-22}\end{equation}

\noindent with $k_{a},$ $k_{d}$ replaced by $k_{a}^{'}$ and $k_{d}^{'}.$
The fraction of blue $+$ blue-green cells is computed from an expression
similar to 22 but with the integration limits ($0.1,$ $0.9$) in
the numerator replaced by ($0.1,$ $1.0$). The calculated dose-response
curve gives a good fit to the experimental data points (solid triangles
in figure 4). The two curves in figure 4 have been obtained for the
same set of parameter values using Mathematica. The good quantitative
agreement between our theoretical results and experimental data indicates
that the stochastic model of GE considered by us captures the essential
features of stochastically induced binary response in GE. 

\begin{figure}[H]
\begin{center}\includegraphics[%
  width=3in]{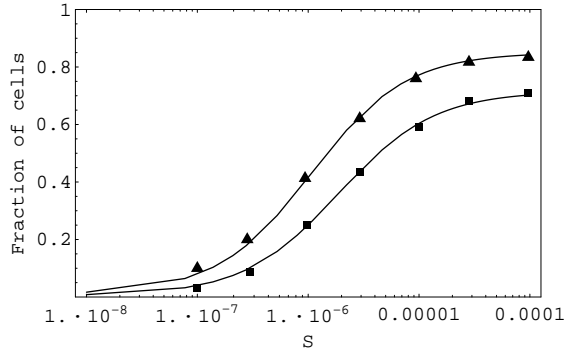}\end{center}

\caption{Fractions of blue $+$ blue-green (upper curve) and blue (lower curve)
cells versus $S$ in a semi-logarithm plot. The experimental points
are depicted by solid triangles signs and solid squares. The parameter
values used for the fitting are mentioned in the text. }
\end{figure}

The reaction scheme 3 doesnot fulfill detailed balance, i.e, equilibrium
conditions. The steady state concentrations in equation (5) are derived
from the more general non-equilibrium conditions of stationarity.
Ref. \cite{key-36} provides examples of reaction schemes related
to GE which violate detailed balance conditions. One of these involves
the phenomenon of stochastic focussing (SF) in which signal fluctuations
sharpen the response in a regulated process. SF appears to be an out-of-equilibrium
effect which is absent if the reaction scheme is constrained by detailed
balance. Modifications of reaction scheme 3 which preserve detailed
balance are possible and it will be of interest to determine whether
SBR can occur in both equilibrium and out-of-equilibrium scenarios.

\begin{flushleft}\textbf{\Large 4. Conclusion and outlook}\end{flushleft}{\Large \par}

\noindent In this paper, we have considered a simple stochastic model
of GE and demonstrated that stochasticity provides the basis for graded
and binary responses to inducing signals. The sole ingredients of
the minimal model of stochastic GE studied in this paper are gene
activation, deactivation, protein synthesis and degradation, each
of which involves a number of biochemical events. Stochasticity in
this model is associated only with the gene activation and deactivation
processes whereas protein synthesis and degradation are assumed to
occur in a deterministic manner. A deterministic description of protein
synthesis is justified when the number of proteins produced is large.
This is the situation in the experiment by Zlokarnik et al. \cite{key-6}
in which proteins per cell are a few thousands in number. For smaller
protein numbers, the inclusion of stochasticity during protein synthesis
and degradation is expected to blur the GE responses but the major
conclusions of the paper still remain valid. The processes of transcription
and translation in the model are not treated separately but lumped
together in a single protein synthesis step. In an eukaryotic cell,
combining transcription and translation into a single step may be
considered to be a drastic approximation. One can study the effect
of stochastic gene activation and deactivation on the transcription
process itself and focus on $mRNA$ synthesis rather than proteins
in reaction schemes \ref{mathed:first-eqn}-\ref{mathed: eqn-3}.
This type of approach highlights the quantal nature of transcription
with bursts of $mRNAs$ being produced in a probabilistic manner in
agreement with experimental observations \cite{key-9,key-30}. In
fact, the value $k_{p}=1$ is more appropriate if $k_{p}$ is interpreted
as the $mRNA,$ rather than protein decay constant. The mathematical
analysis and conclusions are the same as before since protein production
is linked to $mRNA$ synthesis. Despite the limitations of the model,
it contains the important features necessary for an explanation of
the stochastically induced {}``all-or-none'' phenomenon observed
in some eukaryotic systems. The model results give a good description
of the experimental data of Zlokarnik et al. \cite{key-6} and are
expected to be of relevance in explaining the binary response in GE
observed in other eukaryotic systems \cite{key-4,key-5}. The probabilistic
nature of gene activation and deactivation processes is crucial to
explain how graded and binary responses in GE occur in the model.
The stochastic origin of binary response is distinctive from the binary
response brought about by positive feedback processes. Experiments
designed to probe the stochastic origins of graded and binary responsPlot
of $p(x)$ versus $x$ (Case I, $r_{1},\: r_{2}>1$): 1A ($r_{2}>>r_{1})$,
1B ($r_{2}=r_{1})$ and 1C ($r_{1}>>r_{2})$ respectively.es, are
needed for a clearer understanding of the role of stochasticity in
such responses. The stochastic model of GE,corresponding to reaction
scheme 2, has earlier been studied to explore the stochastic origins
of haploinsufficiency \cite{key-13,key-17,key-37}. The model studied
in the paper is a modification of the earlier model. The simplicity
of the models allows for mathematical analysis and helps in identifying
the origins of phenomena associated with stochastic GE. The knowledge
and insight gained from the study of simple models like the present
one provide necessary inputs to develop more detailed and realistic
models of GE.

\begin{center}\textbf{ACKNOWLEDGEMENT}\end{center}

\noindent The Authors thank Gregor Zlokarnik and David Hume for helpful
comments. R. K. is supported by the Council of Scientific and Industrial
Research, India under Sanction No. 9/15 (239) / 2002 - EMR - 1.

\end{document}